\documentclass{iau} 
\usepackage{graphicx}
\usepackage{amsmath}

\title{Using MUSE-AO observations to constrain the formation of the large nuclear star cluster in FCC\,47}

\author[Katja Fahrion \& Mariya Lyubenova \& Glenn van de Ven \& Michael Hilker]   %% give here short author list %%
{Katja Fahrion$^1$
 \and Mariya Lyubenova$^1$
 \and Glenn van de Ven$^2$
 \and Michael Hilker$^1$}

\affiliation{$^1$European Southern Observatory, \\ Karl-Scharzschild-Stra\ss{}e 2, 85748 Garching bei M\"unchen, Germany \\ email: {\tt kfahrion@eso.org} \\[\affilskip]
$^2$Department of Astrophysics, University of Vienna, \\ T\"urkenschanzenstrasse 17, 1180 Wien, Austria \\}

\pubyear{2019}
\volume{351}  
\jname{Star Clusters: From the Milky Way to the Early Universe}
\editors{A. Bragaglia, M.B. Davies, A. Sills \& E. Vesperini, eds.}
\begin{document}

\maketitle
%. CONTINUE EDITING FROM HERE

\begin{abstract}
Nuclear star clusters (NSCs) are found in at least 70\% of all galaxies, but their formation path is still unclear. In the most common scenarios, NSCs form in-situ from the galaxy’s central gas reservoir, through merging of globular clusters (GCs), or through a combination of the two. As the scenarios pose different expectations for angular momentum and stellar population properties of the NSC in comparison to the host galaxy and the GC system, it is necessary to characterise the stellar light, NSC, and GCs simultaneously. Wide-field observations with modern integral field units such as the Multi Unit Spectroscopic Explorer (MUSE) allow to perform such studies. However, at large distances, NSCs usually are not resolved in MUSE observations. The particularly large NSC ($R_\mathrm{eff} \sim$ 66 pc) of the early-type galaxy FCC\,47 at distance of $\sim$ 20 Mpc is an exception and is therefore an ideal laboratory to constrain NSC formation of external galaxies. 

\end{abstract}

\firstsection % if your document starts with a section,
              
\section{Introduction}
Nuclear star clusters (NSCs) are extremely dense stellar systems that reside in the photometric and kinematic centres of their host galaxies (\cite{Boker2002}, \cite{Neumayer2011}). They have similar sizes to globular clusters (GCs), but often are more massive and thus denser. It has been found that NSCs can co-exist with central super massive black holes (SMBHs, e.g. \cite{Georgiev2016}) with the Milky Way being the most prominent example (\cite{Feldmeier2014}). Studies have shown that they follow the same relations with their host's properties as SMBHs, but NSCs allow to extend these relations to lower masses of the central object (e.g. \cite{OrdenesBriceno2018}). These relations between NSC and host can indicate a connected evolution, however, the formation of NSCs themselves is still debated. 

The currently discussed scenarios propose that NSCs could either form from infalling gas directly at the galactic center (e.g. \cite{Bekki2006}, \cite{Antonini2015}) or through the mostly gas-free accretion of GCs that spiral inwards due to dynamical friction (e.g. \cite{Tremaine1975}; \cite{ArcaSedda2014}), but also a composite scenario is possible that involves the merger of young, gas rich clusters at the galactic centre (\cite{Guillard2016}). In the in-situ scenario, the NSC forms independent of the GC system and due to gas accretion from a disk, a strong rotation, high metallicities, and an extended star formation history are expected.
In the dry GC accretion scenario, the NSC is expected to show less rotation because of accretion from different directions, but simulations have shown that NSCs via GC accretion can have significant rotation (\cite{Hartmann2011}; \cite{Tsatsi2017}), and if young or gas-rich GCs are accreted, the NSC can also show a complex star formation history (\cite{Antonini2014}; \cite{Guillard2016}).

Constraining NSC formation is therefore a convoluted problem that requires a complete view of both the kinematics and chemical properties of the host galaxy, the NSC, and the GC system. Observationally, such a study can be achieved with modern wide-field integral field unit instruments as they allow a simultaneous observation of a galaxy from its centre to the halo and are able to spatially and spectrally resolve the different components. The Multi Unit Spectroscopic Explorer (MUSE) at the Very Large Telescope is a particularly useful instrument for this purpose because of its wide field-of-view of 1$^{\prime} \times 1^\prime$ at a spatial sampling of 0.2$^{\prime\prime} \mathrm{pix}^{-1}$ in combination with its optical wavelength range from 4500 to 9300 \AA\, at 2.5 \AA\,resolution. 

The early-type galaxy FCC\,47 (NGC\,1336) in the outskirts of the Fornax galaxy cluster at a distance of 18.3 Mpc (\cite{Blakeslee2009}) is an excellent laboratory to study NSC formation with MUSE because of its rich GC system with $\sim$250 members (\cite{Jordan2015}, \cite{Liu2019}) and its known large NSC with $R_{\text{eff}} = 0.7^{\prime\prime} = 66$ pc that allows to resolve the NSC spatially. We presented a detailed study of FCC\,47, its NSCs, and the GC system with MUSE in \cite[Fahrion et al. (2019b)]{Fahrion2019b}. Here we want to highlight the key aspects of our analysis and their implications.

\section{Kinematics and stellar population properties of the different components}
We used science verification (SV) adaptive optics (AO) optimised MUSE data of FCC\,47 taken in the wide field mode (WFM) to gain insights into the kinematics and stellar population properties of FCC\,47's stellar body, the NSC, and the GCs. To achieve this, we first extracted the spectra and then fit them to obtain line-of-sight velocity distribution parameters and stellar population properites such as mean age, metallicity and light element abundances. Details can be found in \cite[Fahrion et al. (2019b)]{Fahrion2019b} and similar approaches are described, for example, in \cite[Pinna et al. (2019)]{Pinna2019} or \cite[Poci et al. (2019)]{Poci2019}.

\begin{figure}
\centering
\includegraphics[width=0.99\textwidth]{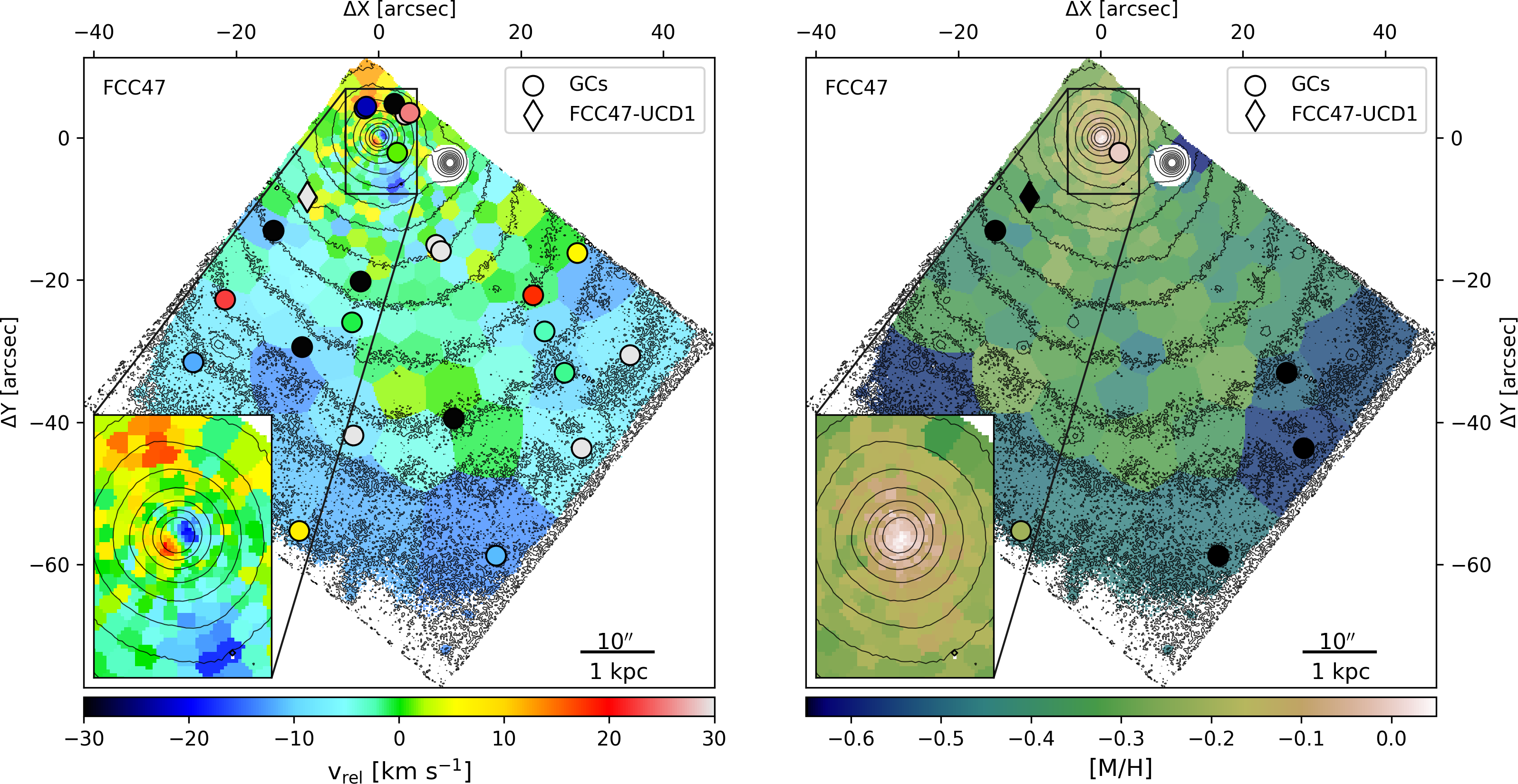}
\caption{Voronoi binned map of the LOS velocity (left) relative to the systemic velocity of 1444.4 km s$^{-1}$ and mean metallicity (right). The circles show GCs that were extracted from the MUSE cube and fitted for their LOS velocity and metallicity, respectively. The diamonds indicate FCC\,47-UCD1 (\cite{Fahrion2019a}). The inset shows a zoom to the central region.}
\label{fig:map_with_gcs}
\end{figure}

After Voronoi-binning of the MUSE cube to ensure a continuous signal-to-noise ratio (S/N) of 100 \AA$^{-1}$ across the cube, the binned spectra were fitted with \textsc{pPXF} (\cite{Cappellari2004}, \cite{Cappellari2017}) using the MILES single stellar population templates (\cite{Vazdekis2010}). This allowed us to create binned maps of the line-of-sight velocity distribution (LOSVD) parameters and stellar population properties. The LOS velocity field of FCC\,47 (left panel in Fig. \ref{fig:map_with_gcs}) shows two kinematically decoupled components (KDCs): the NSC and a rotating disk with a maximum rotation amplitude of $\sim 20$ km s$^{-1}$. The metallicity map shows a confined peak at the location of the NSC and then a steep gradient to lower metallicities. The NSC reaches the oldest ages ($\sim$ 13 Gyr) while the outskirts of the galaxy are younger ($\sim 8$ Gyr, see Fig. 7 in \cite{Fahrion2019b}).

The exceptionally large size allowed to resolve the NSC in the MUSE data. To extract a clean, high S/N spectra for stellar population analysis, we subtracted the galaxy light in an annulus around the NSC (see Fig. 4 in \cite{Fahrion2019b}). The \textsc{pPXF} fit reveals that the NSC is old ($\sim$ 13 Gyr) and metal-rich ([M/H] = 0.08 dex). It is dominated in mass by an old, metal-rich population, but also contains a metal-poor population of $\sim$ 30\% in mass. The total stellar mass of the NSC amounts to $\sim7 \times 10^8 M_\odot$, around 5\% of the total stellar mass of FCC\,47.

The detection of GCs is challenging due to the bright galaxy background, so a model of the surface brightness was subtracted from the collapsed MUSE cube to create a residual image. We then detected GCs after cross-referencing with the ancillary photometric GC catalogue of \cite{Jordan2015}. To extract GC spectra, we used a circular aperture weighted by the point spread function ($\sim$ 0.7$^{\prime\prime}$) and further subtracted the local galaxy background spectra obtained in an annulus placed around each GC. In addition to the detection of a ultra compact dwarf galaxy (FCC\,47-UCD1 \cite{Fahrion2019a}), we determined the LOS velocities of 24 GCs with S/N $> 3$ \AA$^{-1}$\, (circles in the left panel of Fig. \ref{fig:map_with_gcs}) and metallicities of five GCs with S/N $> 10$ \AA$^{-1}$ (right panel). We did not find any rotating substructure within our limited sample of GC LOS velocities, and four out of five GCs have significantly lower metallicities than both the underlying galaxy and the NSC. We found a total mass in the GC system of $\sim1 \times 10^8 M_\odot$. The red, metal-rich GCs are centrally concentrated, but only constitute $\sim$30\% of all GCs (\cite{Jordan2015}). 
 
\section{Implications}
We interpret the results from our analysis of the MUSE + AO WFM SV data of FCC\,47 in the context of NSC formation, focusing on the two most discussed scenarios: the (dry) GC accretion scenario and the in-situ formation.

{\underline{\it GC accretion scenario}}. 
$N$-body simulations have shown that the angular moment of FCC\,47's NSC can not be reached with the random infall of GCs, but requires a preferred orbital infall direction (see \cite{Lyubenova2019} and \cite{Tsatsi2017} for details).
The high metallicity and the old age of the NSC imply that it must have formed dominantly from metal-rich GCs, maybe similar to the red, centrally GC subpopulation that is still present in the galaxy today. With our limited sample, we cannot detect any rotating substructure that might be present in these red, metal-rich GCs and further studies with more LOS GC velocities are required. 

With its rich GC system, FCC\,47 is already an outlier compared to other galaxies of its mass (\cite{Liu2019}) and just from simple dynamical arguments follows that a number of GCs must have fallen into the NSC during the evolution of the galaxy. However, it is questionable if the high mass of the NSC can be explained solely by GC accretion. Our estimate shows that the NSC is several times more massive the total GC system observed today. To explain its high mass and high metallicity with the accretion of gas-free GCs, it must have accreted hundreds of massive, already metal-rich GCs early on.

{\underline{\it In-situ formation scenario}}. 
Besides the infall of GCs from a preferred direction, the angular momentum of the NSC can also be explained naturally in the in-situ formation scenario where the rotation arises from angular momentum conservation of gas that settles at the galactic centre. At the same time, the high mass and metallicity of the NSC can also be explained in this scenario, assuming a large amount of gas was funnelled to the centre initially and underwent fast self-enrichment. The absence of a young population in the NSC is quite striking and signifies that no significant amount of gas was supplied to the NSC later on. We do not now how the star formation in the NSC was quenched, but feedback from strong star formation or an active galactic nucleus could be responsible.

{\underline{\it Evidence for merger history}}. 
The presence of KDCs is usually associated with major mergers. Many studies have been successful to create stable KDCs in simulations of galaxy mergers, both in gas-rich and gas-free mergers (e.g. \cite{Tsatsi2015}, \cite{Rantala2019}). Thus, the presence of two KDCs in FCC\,47's velocity field, one of them the NSC itself, likely indicates at least one major merger in the galaxy's past that has altered the kinematic structure of the galaxy significantly.
The merger history of FCC\,47 might not only be important for the formation of the galaxy itself, but also for its NSC. Studies of other massive NSCs and their host galaxies are required to reveal the role of galaxy mergers in forming the complex galaxy-NSC system.

\end{document}